\documentstyle[12pt,psfig]{article}

\newcommand{\doublespace}{
\renewcommand{\baselinestretch}{1.6}\large\normalsize}
 
\begin{document}

\hyphenation{plaq-uette}

\begin{titlepage}

\begin{tabbing}
\`hep-lat/9601019\\
\`SWAT/96/97  \\
\` January, l996 \\
\end{tabbing}
 
\vspace*{1.0in}
 
\begin{center}
{\bf Confinement by Monopoles in the Positive Plaquette Model of 
SU(2) Lattice Gauge Theory\\}
\vspace*{.5in}
John D. Stack \\
\vspace*{.2in}
{\it Department of Physics, \\
University of Wales, Swansea, \\
Singleton Park,\\
Swansea SA2 8PP, U.K.\\
and\\
 Department of Physics,$^{*}$ \\
University of Illinois at Urbana-Champaign, \\
1110 W. Green Street, \\
Urbana, IL 61801, U.S.A. \\}
\vspace*{.2in}
Steven D. Neiman\\
\vspace*{.2in}
{\it  Department of Physics, \\
 University of Illinois at Urbana-Champaign, \\
1110 W. Green Street, \\
Urbana, IL 61801, U.S.A. \\}
\vspace*{.4in}
\end{center}

\begin{tabbing}
\` PACS Indices: 11.15.Ha\\
\` 12.38.Gc\\
\end{tabbing}
$^{*}$permanent address

\end{titlepage}
\vfill\eject

\doublespace
\pagestyle{empty}

\begin{center}
{\bf Abstract}
\end{center}

  Confinement via 't Hooft-Mandelstam monopoles is studied for the positive
  plaquette model in $SU(2)$ lattice gauge theory.  Positive plaquette model
  configurations are projected into the maximum abelian gauge and the magnetic
  current extracted.  The resulting magnetic current is used to compute
  monopole contributions to Wilson loops and extract a monopole contribution
  to the string tension.  As was previously found for the Wilson action,
  the monopole contribution to the string tension agrees with the string 
  tension calculated directly from the $SU(2)$ links.  The fact that the 
  positive plaquette model suppresses $Z_2$ monopoles and 
  vortices is discussed.

\vspace*{.5in}
 
\newpage
\pagestyle{plain}

	This paper contains our results on confinement by monopoles 
	in pure SU(2) lattice gauge theory, for an action known as the positive
	 plaquette model (PPM).  This work is part of our ongoing 
	lattice gauge theory investigation of confinement by monopoles.  
	In our previous work we have obtained quantitative results for the
	string tension using monopoles 
	in $U(1)$ lattice gauge theory for both $d=3$ and $d=4$, 
	\cite{ws,ws2,ws3}
	and for $SU(2)$ using
	the standard Wilson action (WA), in $d=4$ \cite{jssnrw}.

	The action for the PPM agrees with
	that of Wilson except that the  PPM completely suppresses plaquettes
	with negative trace, {\it i.e.} negative plaquettes are regarded
	as having infinite action.
	In the weak coupling or small lattice spacing limit, 
	the two actions are equivalent. 
	For calculations at finite lattice spacing $a$, the PPM
	represents an ``improved" action.  A plaquette with negative 
	trace is a clear
	lattice artifact.  
	Writing the usual product of links around a plaquette as
	$U_{P}=\exp(ia^{2}F_{\mu\nu})$, a negative
	plaquette has $tr(U_{P})<0$, which corresponds to a field-strength
	$F_{\mu\nu} \sim O( \pi/a^2)$.  
	Such large field strengths play no role in
	the continuum limit, and their suppression in the PPM
	should allow a clearer view of the continuum limit to be obtained from
	calculations performed at finite lattice spacing $a$.  Recently the
	PPM has been subjected to a thorough study \cite{fingberg}, 
	which shows it to be
	in the same universality class as the Wilson action.  
	However, while the Wilson action possesses a well-known
	dip in the step $\beta$-function,~$\Delta\beta(\beta)$,  no such
	dip occurs for the PPM.

	There is a more 
	specific reason for exploring monopole confinement in the PPM.  
	The Wilson action, since it permits negative plaquettes, contains
	$Z_2$ monopoles and vortices.  ( $Z_2$ monopoles are associated with
	cubes whose faces contain an odd number of negative plaquettes, 
	and $Z_2$
	vortices similarly require the presence of negative plaquettes.)
	These $Z_2$ objects are associated with the center of the $SU(2)$
	group, and 
	there is a long history of attempts to understand confinement in
	$SU(2)$ using them \cite{tomb,mack1,mack2,mack3,mack4}.  
	However, for the PPM, there are no negative plaquettes, 
	and therefore no lower bound on the $SU(2)$ string tension for
	the PPM can be obtained by considering $Z_{2}$ objects.
	(We assume the
	latter are 
	defined using single
	plaquettes.)
	In contrast, the 't Hooft-Mandelstam or
	dual superconductor monopoles\cite{thooft,mand}
	will be shown below to
	give a quantitative explanation 
	of the $SU(2)$ string tension for the PPM.

        A correlation
	length large compared to the lattice spacing is desired for the
	continuum limit, but finite lattice size forces a 
	compromise.  In our previous work with the Wilson action on $16^4$
	lattices, this compromise
	was struck with correlation lengths $\xi=1/\sqrt{\sigma} \sim
	5a$, corresponding to couplings  $\beta_{WA}\sim 2.5$.
	In the PPM, a  $16^4$ lattice was also used.
	To determine the
	values of $\beta_{PPM}$ which correspond to $\xi \sim 5a$, we 
	assumed universality of the ratio $T_{c}/\sqrt{\sigma}$, which is
	known to be $0.69(2)$ for the Wilson action \cite{fingberg2}.  
	The couplings for various
	deconfining temperatures have
	recently been determined very accurately for
	the PPM \cite{fingberg}; in particular, for  $T_{c}=1/8a$, 
	$\beta_{PPM}$=1.886(6).  
	Assuming  the PPM has the same value of
	$T_{c}/\sqrt{\sigma}$ as the Wilson action, 
	we then obtain
	$\xi \sim 5.5a$  for
	$\beta_{PPM}=1.886$.  
	Our runs in the PPM were carried out
	at $\beta_{PPM}$=1.886, and the smaller values 1.840 and 1.790.

	The calculation proceeded in a similar manner to our previous
	work with the Wilson action.  For each coupling mentioned above,
	500 configurations were gathered, where every 20th configuration
	was saved after equilibrating for 1000 sweeps.  An update of the
	lattice consisted of 1 Kennedy-Pendleton sweep \cite{kennedy} 
	and two overrelaxation
	sweeps \cite{creutz}.  Any new links that resulted in a negative
	plaquette were rejected.

	To locate monopoles,
	configurations
	are projected with high accuracy into the 
	maximum abelian gauge.  The gauge-fixed links are factored into
	a ``charged" part, times a $U(1)$ link.  
	The monopole location procedure
	of Toussaint and DeGrand is applied to the $U(1)$ links, 
	and results in an integer-valued magnetic current 
	$m_{\mu}(x)$\cite{degrand}.
	The procedure in effect locates a monopole by finding the end of its
	Dirac string.

\newpage   

	The maximum abelian gauge is attained when 
	
\begin{equation}
X(y)\equiv \sum_{\mu}\left[U_{\mu}(y)\sigma_{3}U^{\dagger}_{\mu}(y)
+U^{\dagger}_{\mu}(y-\hat{\mu})\sigma_{3}U(y-\hat{\mu})\right]
\label{Xdef}
\end{equation}
	is diagonal \cite{jssnrw}.  Perfect diagonalization of $X(y)$ is 
	never achieved.
	An overrelaxation process is used to repeatedly sweep the lattice,
	stopping when the off-diagonal elements of $X$ are sufficiently small.
	We used 
\begin{equation}
\left<|X^{ch}|^{2}\right>\equiv \frac{1}{ L^{4}}\sum_{x}\left(
|X^{1}(x)|^{2}+|X^{2}(x)|^{2}\right)
\end{equation}
	as a measure of the off-diagonal elements of $X$, where $X_{1},X_{2}$
	are the coefficients of $\sigma_{1},\sigma_{2}$ in a Pauli matrix
	expansion of $X$.  The
	overrelaxation process was stopped when  
	$\left<|X^{ch}|^{2}\right>\le 10^{-10}$.  
	This required approximately $1000$ overrelaxation sweeps.  

	We now have two methods of 
	extracting a potential; one directly from the $SU(2)$ links, the other
	using the magnetic current to calculate monopole Wilson loops and
	a monopole potential \cite{jssnrw}.  
	In both cases, all Wilson loops  $W(R,T)$ 
	were measured
	up to a maximum size of $8\times 12$.  The full $SU(2)$ and monopole
	potentials were determined by linear fits of $\ln(W(R,T)$ 
	vs $T$.  These fits
	were done for $R \geq 2$ over the interval $T=R+1$ to $12$.  Having
	determined  potentials $V(R)$ for each $R$, the string tensions 
	$\sigma$ were found
	by fitting $V(R)$ to the form $V(R)=\alpha/R+\sigma\cdot R+V_{0}$,
	over the interval $R=2$ to $R=8$.  The full $SU(2)$ and monopole 
	determinations of the string tension are shown in Table I, 
	where it can
	be seen that the two are in excellent agreement.  The agreement is of
	the same quality as in our previous work with the Wilson action, and
	is another piece of evidence in favor of confinement via 
	't Hooft-Mandelstam monopoles.  
                                                                                
	The PPM monopole potentials are shown in 
	Figure 1, where the solid curves are the results of the 
	linear-plus-Coulomb fits.  The Coulomb coefficients $\alpha$ naturally
	differ for the monopole and full $SU(2)$ potentials. 
	The monopole contribution
	is purely non-perturbative and is supposed 
	to be correct only in the large distance
	region.  
	In particular the monopole potentials do  not contain the Coulombic
	term coming from one gluon exchange
	present in the full $SU(2)$ potential.
	For the three couplings,
	$\beta_{PPM}=1.886,1.840$, and $1.790$, the values of $\alpha$ from the
	monopole potentials are very small; $0.01(1),0.02(1),0.02(1)$ 
	respectively.  For the full $SU(2)$ PPM potentials the corresponding
	results for $\alpha$ are $-0.29(1),-0.30(1),-0.31(1)$.  Both of these
	sets of results for $\alpha$ in the PPM are very similar to those
	obtained previously for the Wilson action.

\begin{table}
\begin{center}
\begin{tabular}{||c|c|c||} \hline
$\beta_{PPM}$ & $\sigma_{SU(2)}$ & $\sigma_{mon}$\\ \hline
1.790 & 0.041(1) & 0.043(2) \\ \hline
1.840 & 0.036(1) & 0.036(1) \\ \hline
1.886 & 0.029(1) & 0.028(2) \\ \hline
\end{tabular}
\end{center}
\begin{center}
\bf{Table I}
\caption {The string tensions from monopoles and full
$SU(2)$ for the PPM}
\end{center}
\end{table}

	The distributions of magnetic current for the Wilson action and positive
	plaquette models are qualitatively similar, but there are 
	interesting quantitative differences.  For the PPM coupling 
	$\beta_{PPM}=1.840$, the string tension is $0.036(1)$; within error bars
	of the string tension for the Wilson action at $\beta_{WA}=2.50$, 
	where our
	result was $0.034(1)$.  These two couplings are thus approximately 
	equivalent in terms of the physical string tension they produce.  
	For the Wilson action at $\beta_{WA}=2.50$, the 
	fraction of links carrying
	magnetic current is
	$1.36(1)\times 10^{-2}$, whereas for the PPM coupling 
	$\beta_{PPM}=1.840$,
	the corresponding number is significantly 
	smaller,$\ 1.12(1)\times 10^{-2}$.  In terms of
	the number of links with current, there are approximately 600
	more links of magnetic current for the case of the
	$\beta_{WA}=2.50$ Wilson action 
	on a $16^4$ lattice.  
	The nature of the difference becomes clear when the
	magnetic current is resolved into individual loops, each satisfying
	current conservation.  For either action, a substantial fraction of
	the current 
	resides in small loops of 4,6, {\it etc.} links.  
	These small loops of current
	have nothing to do with confinement.  
	Rather it is the large loops with
	numbers of links ranging from 50 up to several hundred which are 
	responsible for the string tension.  Specifically, 
	if loops of size less than 50 links
	are 
	eliminated from the magnetic current, the physical string tension is
	reproduced for both the $\beta_{WA}=2.50$ Wilson action and the PPM at
	$\beta_{PPM}=1.840$.  Eliminating  the contribution 
	to the current 	coming from loops of less than 50 links, 
	the fraction of lattice links
	occupied is 
	$0.60(1)\times 10^{-2}$ for the Wilson action, very close to
	$0.58(1)\times 10^{-2}$, which is the result for the PPM action.   
	So for that part of the
	magnetic current which is effective in producing the string tension,
	namely loops of current of 50 links and larger,
	the two actions have essentially the same fraction of links with
	current.  The excess found for the Wilson action involves 
	small loops
	of magnetic current, which play no role in confinement. This suggests
	that for successively improved actions, the fraction of magnetic 
	current in small loops will steadily fall.

	To summarize, we have demonstrated for SU(2) lattice gauge theory, 
	that the quantitative explanation
	of confinement  by
	't Hooft-Mandelstam monopoles is the same for the PPM as it was for
	the standard Wilson action.  This shows that the picture is robust.
	Further, the results for the PPM make clear that $Z_2$ monopoles
	and vortices are not responsible for the string tension.

This work was supported in part by the National Science Foundation under
Grant No. NSF PHY 94-12556, and the Higher Education Funding Council for Wales
(HEFCW).  The calculations were carried out on the Cray
C90 system at the San Diego Supercomputer Center (SDSC),
supported in part by the National Science
Foundation.

\newpage
\begin{figure}
\psfig{file=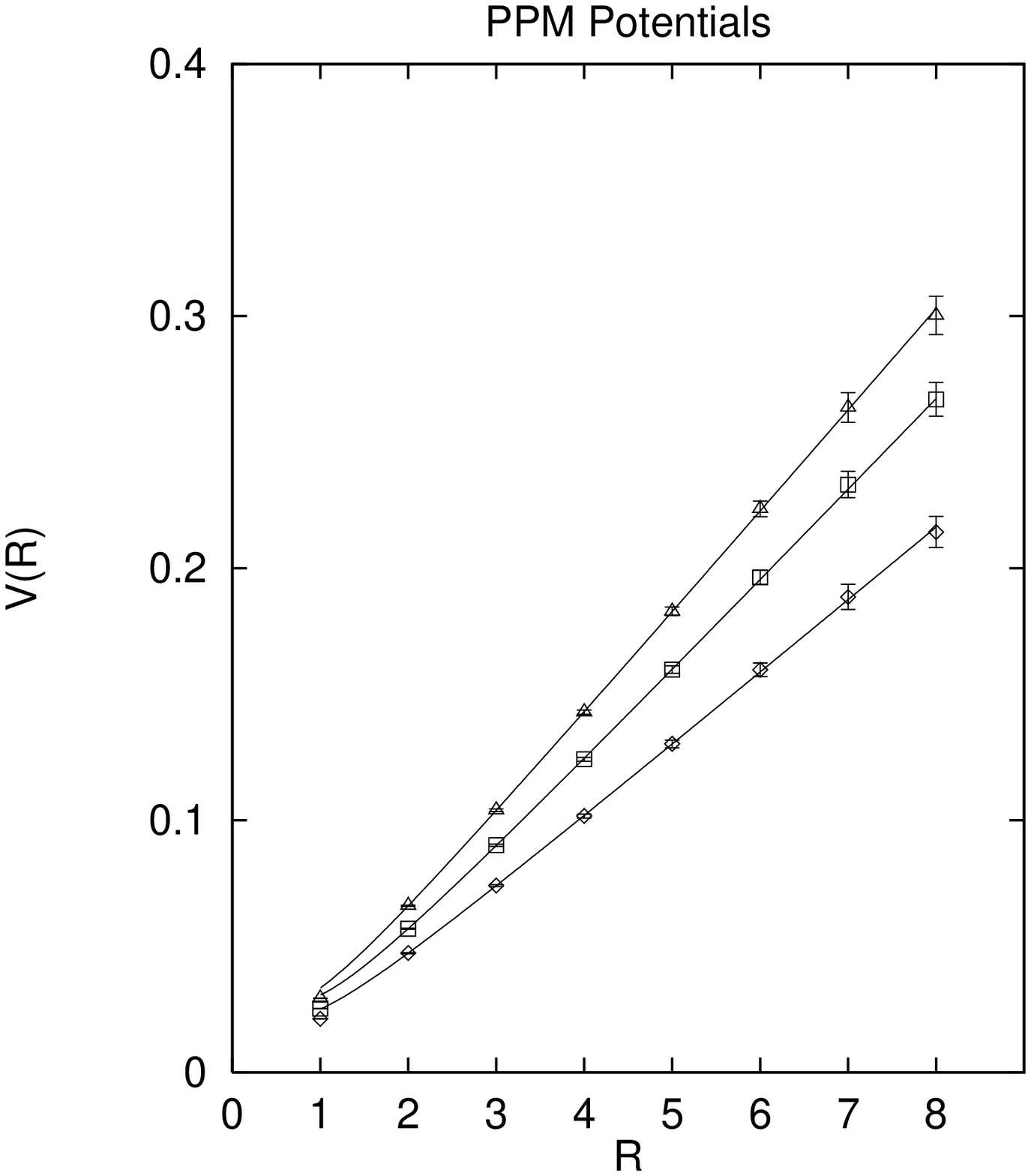}
\caption{  The potentials extracted from monopole Wilson
loops for the PPM at $\beta_{PPM}=1.790$ (triangles), 1.840 (squares), 
and 1.886 (diamonds).
The solid lines are the linear-plus-Coulomb fits to each potential.}
\end{figure}

\end{document}